\documentclass[12pt]{article}
\usepackage{epsfig}
\usepackage{amsmath}
\begin{document}

\begin{center}
{\Large \bf Theoretical and real-time study of uniaxial nematic liquid crystal phase transitions using Fresnel diffraction}
\vspace{5mm} \\{\normalsize N. Madadi and M. Amiri $^{*}$} \\
\vspace{2mm} {\normalsize \it Department of Physics, Bu-Ali
Sina University,\\Hamedan 65175-4161, Iran}
\end{center}
%
%---------------------------------------------------------------------------------------------------------------------
%
\begin{abstract}
Liquid crystals (LCs) play a fundamental and significant role in modern technology. Recently, they have also been used in active switching, adaptive optics, and next-generation displays for augmented and virtual reality. This is due to the diverse properties of their various phases and the growing physical understanding of LCs. Our goal is to examine the applicability of a new method in determining these quantities for thermotropic uniaxial nematic liquid crystals (NLCs), even though nearly all theoretical and experimental efforts are focused on a deeper understanding of the temperature-dependent free energy behavior and other quantities related to it, especially in the vicinity of the first- and second-order phase transitions of LCs. The method that is being discussed is based on Fresnel diffraction (FD) from phase objects, which has found a wide range of precise metrological applications over the past two decades. Diffractometry is a very sensitive, accurate, and immune technique that can convert any change in the order of LCs as a function of temperature into a change in the optical phase and, as a result, a recordable change in the visibility of the light diffraction pattern from phase steps. This contrasts with interferometry, which is very sensitive to environmental changes. Theoretical investigations, numerical calculations, and comparisons of the results with experimental observations in turn demonstrate a very high compliance with the output of other existing methods. As we will see, this method has the potential to not only strengthen existing approaches by addressing some of their flaws and shortcomings but also to take its place next to them.
\end{abstract}
%keywords: Uniaxial nematic liquid crystals, Fresnel diffraction, Diffractometry, Phase transitions, Accurate metrology, Real-time recording
%OCIS codes: 
%PACS number(s):
% %---------------------------------------------------------------------------------------------------------------------
%
%\begin{center}
\section{Introduction}
%\end{center}
Physicists encountered some reversible and puzzling new evidence of organic materials exhibiting mesomorphic phases as a function of temperature between perfectly 3D organized crystals and isotropic liquids in the latter half of the nineteenth century. The first instance of this kind of LC was identified in 1888 \cite{Rei88,Leh89}. The macroscale property of birefrengence is manifested in the intermediate mesophase of anisotropically organized thermotropic LCs \cite{OswPie05}. The rigorous and thorough foundation for defining LCs as the fundamental motive in liquid crystal display (LCD) technology that can be employed at room temperature was largely made possible by Heilmeier's successful and ground-breaking research in the RCA laboratory starting in 1964 \cite{Cas05,Kel73,Kaw02}. Since the inception of LC chemistry, organic compounds having flat series of two or more rings as the core of an enlongated molecular structure with high stiffness compared to the flexible terminal tails have been frequently utilized in electronic displays \cite{DemVil98}. Due to the widespread use of LCs, the topic has been studied and developed steadily, giving rise to various studies and cutting-edge applications. It continues to be one of the most challenging research fields. Investigations have been made into the phase transition behaviors of LCs \cite{FanSte70,GelBar77,Sin00,RaiPas06}, wide-band or selective light reflection \cite{MitSop99,RelMit06,LuChi07}, selective optical properties and controllable ferroeletric LCs \cite{HikKem98,MeyKel75,GarMey78,AndSte87}, fluctuation effects at phase transitions \cite{Sin00}, the effect of applied fields \cite{Wil63,SteStr74}, high resolution adiabatic scanning calorimetry and ac calorimetry technologies \cite{ThoGlo09,CorTho10}, and their function in next-generation displays \cite{CheWu18,ErsMul20,XioWu21} have all been investigated. A wide variety of very valuable optical devices, particularly in imaging and adaptive optics, have emerged as a result of recent advancements in continuous tunability LCs, which are significantly dependent on international competition and cooperation \cite{Vic03}.\\
In this work, we present a simpler method for analyzing phase transitions in thermotropic LCs based on unconventional FD from a phase step. As is well known in classical wave optics, when a portion of the wave-front of a spatial coherent light wave is blocked by an opaque object, the light intensity is redistributed over the diffraction pattern right next to the geometric shadow \cite{BorWol99}. In other words, it will be assumed that an actual source for the development of FD was the existence of a singularity in the amplitude of the wave-front. But unlike the standard FD, the most advantageous feature of the unconventional FD is its tunability, which appears in connection with one of the various coherent light wave singularities listed below:\\
i) When a coherent light wave propagates parallel to the plane interface between two media in transmission and passes through two adjacent isotropic plane parallel transparent plates with sharp, not wavy, or uneven edges and different indices of refraction that are set to the same thickness \footnote{The analytical complexity of the optical path difference and shadowing correction will therefore be reduced in some way in the transmission and reflection of a monohromatic light wave.}; or when a reflector step with a variable height is illuminated at normal incidence by a coherent light wave in reflection (i.e. singularity in the optical path or dislocation in the phase of a wave-front) \cite{AmiTav07}.\\
ii) After leaving a phase object, a coherent light beam can split into two crossing wave-fronts \cite{TorHos18}. For instance, if a coherent and collimated light beam traverses a Fresnel mirror or a Fresnel biprism (i.e. singularity in wave propagation direction or dislocation in wave-front gradients).\\	
iii) Similarly, it is found that the FD fringes are produced when the optical field's polarization and coherency are rapidly changed.\\\\
The resulting wave-front dislocations as well as the gradient created by the phase objects can be changed at will in the typical situations mentioned above. The most significant implication is that it may be used to transform any change in the physical properties of transparent substances into a phase change, which alters the visibility and location of diffraction fringes. Conversely, by carefully examining the intensity distribution and/or the extrema positions of the periodic diffraction fringes, it is possible to accurately determine the refractive index, a crucial characteristic parameter of a substance, as well as the characterization of the polarization of light rays in the study of optical transmission in any transparent medium.\\
The first systematic and in-depth study of diffraction from a phase step was published by M. T. Tavassoly {\it et al.} in the early 2000s \cite{TavHos01}, and it has since expanded into a variety of technological fields over the past two decades \cite{TavMor09}. Because of this, it has led to the development of numerous new optical metrology features and applications, such as the measurement of film thickness \cite{TavHas09}, high-precision refractometry \cite{TavSab10,TavHas12}, nanometer displacement measurement \cite{KhoTav12}, refractive index gradient in inhomogeneous phase objects and diffusion processes \cite{TavMor09,BeyDas15}, spectral modifications \cite{TavKha05,AmiTav08,AmiAli10}, wavemetry \cite{HosTav15}, real-time monitoring and measurement of chemical etching rate of transparent materials \cite{Mah15}, and 3D imaging \cite{EbrDas19,SiaMor21}, alongside a lot of many other uses inspired by phase objects' FD capabilities.\\ 
Due to LCs' widespread use in modern technology and their birefringent characteristics, the refractive indices associated with these characteristics play a crucial role in the design of instruments, including LCs. Based on quantitative measurements of the periodic diffraction fringes, we aimed to provide a novel description of the temperature dependence of the average refractive index value in nematic LCs as a macroscopic observable at a particular wavelength. A single-shot imaging can also be utilized to determine the number of phase transitions between 3D ordered crystals and isotropic liquids, as well as surface ordering and anchoring behavior of LC molecules, as will be shown by applying this potent technology.
%
%-------------------------------------------------------------------------------------------------------------------
%
%\begin{center}
\section{Outline of principles and description of the method}
%\end{center}
Anisotropic rod-like thermotropic NLCs exhibit short-range order at high temperatures for two reasons: one is due to intermolecular forces, and the other is owing to thermal fluctuations. They maintain their invariance under arbitrary translations and rotations as a result of their perfect symmetry \cite{deGPro93}. As a result of complete symmetry, they remain invariant under arbitrary translations and rotations. To put it another way, the $T(3)\times SO(3)$ symmetry and 3D space symmetry, which characterize the isotropic LC phase, are distinguishing features. On the other hand, solid crystals' long-range positional and orientational order would lead to significantly less symmetry. Though this is not always the case, the first LC phase to appear after cooling down from an isotropic liquid phase is the nematic phase, which has the lower order and higher symmetry. In this phase, a uniaxial/biaxial symmetry destroys the orientational symmetry while maintaining the translational symmetry. As one cools down, the phase transitions can be described in terms of increasing order parameters. A symmetric traceless second-rank alignment tensor can be utilized to calculate the properties of the first-order nematic-isotropic transition using the Landau-de Gennes continuum theory \cite{deGPro93}
%equation (1)
\begin{equation}
Q=S({\bf{\hat{n}}\bf{\hat{n}}}-\frac{1}{3}{\bf I})+P({\bf{\hat{m}}\bf{\hat{m}}}-{\bf{\hat{l}}\bf{\hat{l}}}),
\end{equation}
where ${\bf I}$ denotes the identity tensor, $S$ as the microscopic order parameter is the degree of uniaxial alignment associated with the nematic director $\bf{\hat{n}}$, and $P$ stands for the degree of biaxial alignment associated with the two minor eigen vectors $\bf{\hat{m}}$ and $\bf{\hat{l}}$. In uniaxial NLCs, there are\\
i) The most symmetric phase ($S=0$, $P=0$, and $Q=0$) is the isotropic phase,\\
ii) The less symmetric phase ($S=1$, $P=0$, and $Q=1$) corresponds to a fully aligned nematic phase,\\
iii) The nematic phase is defined as ($0< S <1$ and $P=0$), and $S$ is a scalar order parameter that has the following definition\cite{SauMai61}\\
%equation (2)
\begin{eqnarray}
S=\int^{\pi}_{0}f(\theta)[3cos^{2}\theta-1]d\theta={\frac{1}{2}}<3cos^{2}\theta-1>\nonumber\\
=<P_{_{2}}(\cos\theta)>,
\end{eqnarray}
where $P_{_{2}}(\cos\theta)$ stands for a second-order Legendre polynomial. The order parameter can be approximated using the Haller approximation equation because the thermotropic LC molecules become increasingly disordered as the temperature increases \cite{Hal75}
%equation (3)
\begin{equation}
S=(1-\frac{T}{T_{_{C}}})^{\beta},
\end{equation}
$T_{_{C}}$ is the temperature at which $S$ from the nematic phase to the transition temperature, $T_{_{N-I}}$, abruptly drops to zero due to symmetry considerations and $\beta$ is a fitting parameter. On the other hand, Vuks offered a semi-empirical approach based on the analysis of experimental data for the investigation of the local field in an anisotropic crystal \cite{Vuk66}. This model performs well in the visible and infrared spectral regions and relates the local field to the isotropic field as follows:
%equation (4)
\begin{equation}
\vec{E}_{loc}=\frac{<n^2>+2}{3}<\vec{E}>,
\end{equation}
in which
%equation (5)
\begin{equation}
<n^2>=\frac{1}{3}(n^2_{x}+n^2_{y}+n^2_{z}),
\end{equation}
where anisotropic LCs media can also use the main refractive indices of a crystalline medium, $n_{x}$, $n_{y}$, and $n_{z}$ \cite{Vuk66}. Eq. (5) demonstrates that for uniaxial NLCs
%equation (6)
\begin{equation}
<n^2>=\frac{1}{3}(n^2_e+2n^2_o).
\end{equation}
A LC medium's anisotropically polarizabilities, $\alpha_{e,o}$, and observed average values of their refractive indices, $n_{e,o}$, along and perpendicular to its optic axis are related by \cite{Vuk66}
%equation (7)
\begin{equation}
\frac{n^2_{e,o}-1}{<n^2>+2}=\frac{4\pi}{3}N\alpha_{e,o},
\end{equation}
which is quite interesting and $N$ is the number of LC molecules per unit volume. The optical birefringence of anisotropic LC, defined as $\Delta n=n_{e}-n_{o}$, and the average refractive index of LC, $<n>$, both depend on temperature at a specific wavelength \cite{LiWu004}
%equation (8)
\begin{eqnarray}
<n(T)>=A-BT,\nonumber\\
\Delta n(T)=(\Delta n)_{_{0}}S,\nonumber\\
=(\Delta n)_{_{0}}(1-\frac{T}{T_{_{C}}})^{\beta},
\end{eqnarray}
As a result, LCs' extraordinary and ordinary refractive indices would then depend on temperature
%equation (9)
\begin{eqnarray}
n_{e}\approx <n(T)>+\frac{2}{3}\Delta n(T)\nonumber\\
=A-BT+\frac{2}{3}(\Delta n)_{_{0}}(1-\frac{T}{T_{_{C}}})^{\beta}\nonumber\\
n_{o}\approx <n(T)>-\frac{1}{3}\Delta n(T)\nonumber\\
=A-BT-\frac{1}{3}(\Delta n)_{_{0}}(1-\frac{T}{T_{_{C}}})^{\beta},
\end{eqnarray}
where $A$ and $B$ can be determined by fitting \cite{LiWu05}, and $(\Delta n)_{_{0}}$ denotes the LC birefringence in the crystalline state.\\
\subsection{Guideline of the method}
Consider as a simple illustration an isotropic plane-parallel transparent plate with an abrupt edge, held in air or a transparent isotropic liquid, to get into greater depth and become familiar with the technique. As seen in Fig. 1(a), a 1D phase step is created in transmission mode by submerging a glass slab in a clear, isotropic liquid.
%Fig. (1)
\begin{figure}
\includegraphics[width=0.30\textwidth]{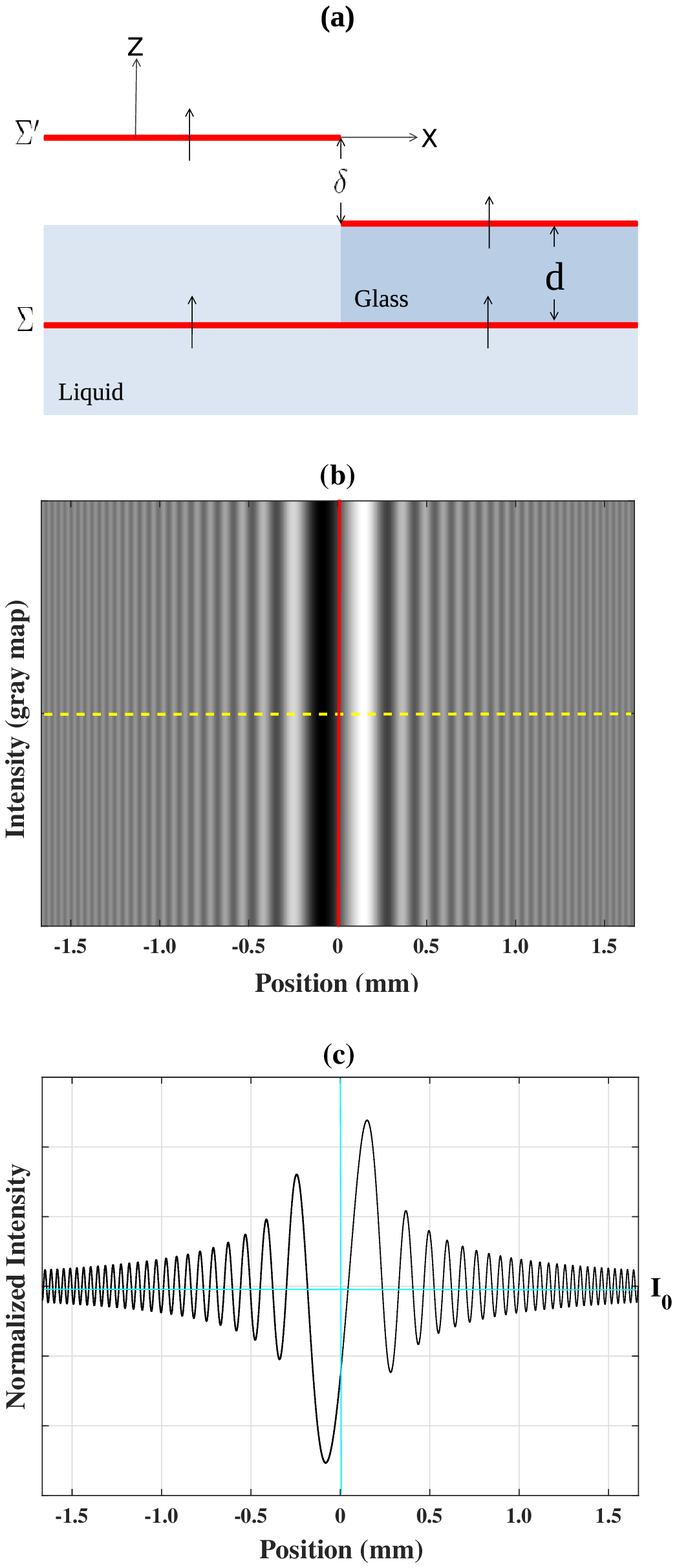}
\caption{(a) The optical length of a direct ray at normal incidence is shown to illustrate the production of a phase step after passing a coherent monochromatic plane wave-front from two adjacent phase objects with different refractive indices. (b) A simulation of the intensity outlook determined at a distance of $R=5$ $cm$ from the back of the glass plate. (c) Calculated normalized intensity distribution of diffracted light along the yellow dashed line, corresponding to (b).}
\end{figure}

The optical path difference (OPD) and phase of the beam rapidly change when a spatially coherent beam of light with propagation direction perpendicular to the transparent material travels through it. Consider a plane wave-front $\Sigma$ that is incident on an abrupt glass edge in the $x > 0$ portion of the $xy$ plane and is propagating in the direction of the $z$-axis. The wave-front $\Sigma '$ that appears at the back of the glass plate, with the left and right segments separated by OPD, $\delta$, can be thought as the source of Huygens-secondary Fresnel's wavelets \cite{AmiTav07}. The height of the phase step is actually represented by the value of $\delta$. Calculating the intensity distribution in the image region parallel to that of the transparent glass can be done using the scalar Kirchhoff diffraction theory. At normal incidence, the phase difference $\phi$ is described as
%equation (10)
\begin{equation}
\phi=k\delta=\frac{2\pi}{\lambda_{_{0}}}(n_{_{glass}}-n_{_{liquid}})d,
\end{equation}
where the refractive indices of the glass, the liquid surrounding it, the vacuum wavelength of the incident light, and the plate thickness are denoted by $n_{_{glass}}$, $n_{_{liquid}}$, $\lambda_{_{0}}$, and $d$, respectively. To accurately measure the refractive index of a LC as a function of temperature for any given wavelength and normal incidence, which is what we would like to do in the current issue, we can similarly use the variation of the diffraction pattern with changing temperature. Prior to addressing this issue, it is important to note how our suggested framework will allow for such.\\
From the diffraction patterns, two techniques are used to calculate the height of the phase step in the wave-front $\Sigma '$ and, subsequently, the refractive index:\\
i) The first technique uses the intensity profile of the diffraction patterns, which is expressed as \cite{AmiTav07} for the selected set of axes
%equation (11)
\begin{eqnarray}
I_{_{L,R}}=I_{_{0}}t_{_{L}}t_{_{R}}[(\frac{1}{2}-C_{_{0}}^{^2}-S_{_{0}}^{^2})\cos\phi\mp(C_{_{0}}-S_{_{0}})\sin\phi]\nonumber\\
+\frac{I_{_0}}{2}[(\frac{1}{2}+C_{_{0}}^{^2}+S_{_{0}}^{^2})(t_{_{L}}^{^2}+t_{_{R}}^{^2})+(C_{_{0}}+S_{_{0}})(t_{_{L}}^{^2}-t_{_{R}}^{^2})],\nonumber\\
=I_{_{0}}t_{_{L}}t_{_{R}}[(\frac{1}{2}+C_{_{0}}^{^2}+S_{_{0}}^{^2})+(\frac{1}{2}-C_{_{0}}^{^2}-S_{_{0}}^{^2})\cos\phi\mp(C_{_{0}}-S_{_{0}})\sin\phi]\nonumber\\
+\frac{I_{_0}}{2}[(\frac{1}{2}+C_{_{0}}^{^2}+S_{_{0}}^{^2})(t_{_{L}}-t_{_{R}})^{^2}+(C_{_{0}}+S_{_{0}})(t_{_{L}}^{^2}-t_{_{R}}^{^2})],
\end{eqnarray}
where $I_{_{0}}$, $t_{_{L}}$, $t_{_{R}}$, $C_{_{0}}$, and $S_{_{0}}$ are the incident light intensity, the left and right side transmission coefficients, and the well-known Fresnel integrals, respectively. Additionally, the $(-)$ and $(+)$ signs, respectively, indicate the equivalent intensities at locations on the left and right sides of the step edge on the wave-front $\Sigma '$ with regard to the coordinate system in Fig. 1(a). According to the diffraction patterns in Figures 1(b) and 1(c), the intensity fluctuation decreases as one moves away from the step edge on both sides and tends to a constant intensity. Quantitatively, in the Fresnel integrals and in the limit $I_{_{L,R}}\rightarrow I_{_{0}}$ as $v_{_{0}}\rightarrow\infty$. As a result, measuring the intensity at a particular position and dividing it by $I_{_{0}}$ will yield the normalized intensity at that point. In general, each arbitrary diffraction pattern have an absolute minimum and two neighboring maxima. Now, it is possible to define the visibility of the intensity pattern's three central fringes as \cite{AmiTav07}
%equation (12)
\begin{equation}
{\cal{V}}=\frac{(I_{_{MaxL}}+I_{_{MaxR}})/2-I_{_{MinC}}}{(I_{_{MaxL}}+I_{_{MaxR}})/2+I_{_{MinC}}},
\end{equation}
where $I_{_{MinC}}$ denotes the utmost minimum intensity on the central dark fringe and $I_{_{MaxL}}$ and $I_{_{MaxR}}$ denote the maximum intensities on both sides of the central dark fringe. The visibility vs normalized OPD function, $\delta/\lambda_{_{0}}=\phi/2\pi$, is periodic, with only one period shown in Fig. 2 as \cite{AmiTav07}
%equation (13)
\begin{equation}
{\cal{V}}=a\sin[b(\frac{\phi}{2\pi})+c],
\end{equation}
where the constants $a$, $b$, and $c$ can be derived from fitting.
%Fig. (2)
\begin{figure}[htb]
\includegraphics[width=0.45\textwidth]{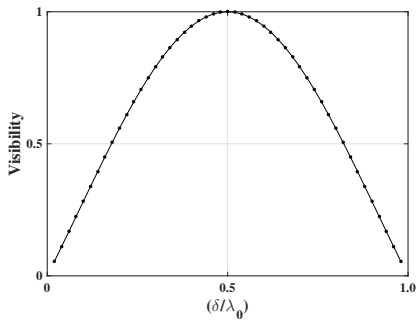}
\caption{Fringe visibility variation with normalized OPD, $(\delta/\lambda_{_{0}})=\phi/2\pi$. For both temperature-dependent linear and non-linear variation of OPD, this function has a periodic and symmetrical structure with an axis of symmetry that has reached its maximum at values of $\phi/2\pi=\pm 1/2, \pm 3/2, . . .$ and reaches zero at values of $\phi/2\pi =0, \pm 1, \pm 2, . . .$.}
\end{figure}
The visibility of the fringes increases from zero to one as normalized OPD varies from zero to one-half, and the value of the visibility and its change are our two key criterion. The maximum visibility occurs in all circumstances where the contributing factors produce phase differences of $\phi=(2m+1)\pi$, for values of $m=0, \pm 1,\pm 2, . . .$. Therefore, for normal incidence at a given wavelength, this number depends solely on the $n_{_{e,o}}$. On the other hand, fitting on experimentally determined visibilities at various temperatures, as a strong approach with a relative uncertainty of around $10^{-5}$ \cite{TavSab10}, gives extensive information on the existing theories about the order parameter as well as the refractive indices of LCs. 
%equation (14)
\begin{equation}
n_{_{e,o}}=n_{_{glass}}\pm\frac{\lambda_{_{0}}}{bd}[\sin^{-1}(\frac{\cal{V}}{a})-c].  
\end{equation}
However, more generally speaking, it should be noted that the refractive index of the glass, $n_{_{glass}}$, does not considerably change with temperature in compared to the surrounding liquid, provided the range of temperatures is not too broad; otherwise, the required corrections must be performed. The sensitivity and resolution of the CCD pixels of the used detector to record data, determine the precision of this method and are of greatest significance in defining the final outcome, even though precision is appropriate to our purpose in our plan. However, the main practical benefit of this approach is that, as long as we are dealing with linear optics, the visibility factor used to calculate the refractive indices is independent of any change in the intensity fluctuations.\\
ii) The second strategy is based on the usage of positions of the diffraction pattern's extrema because changes in any of the factors relating to the physical properties of the 1D transmission phase step would result in a detectable change in the intensity profile of the diffraction patterns. For any given wavelength, as the order parameter changes as a function of temperature, so does the value of $n_{_{e,o}}$ and, as a result, the location of the extrema in the diffraction patterns. By averaging across the positions of the CCD array pixels with the least intensity to cut down on background noise, we can experimentally identify the location of the midpoint of the central dark fringe when the visibility of the diffraction fringes of a 1D phase step is at its highest value. For measuring the positions of the extrema on a CCD array detector, the midpoint should be used as an origin because it coincides with the 1D step edge. However, numerical analyses have demonstrated that, in contrast to all prior situations, the complexity of the behavior of the extremum points in the present study requires that we concentrate on using the visibility of the diffraction fringes in this approach.\\
\section{Numerical results and discussion}
In general, the same techniques used to calculate the refractive indices of bulk crystals may be used to calculate those of uniaxial NLCs. In contrast to crystals, thermotropic LCs also have the property of birefringence, hence it is important to remember this before moving on to the technique that the optical axis cannot be pre-defined. To be able to measure the refractive index in a certain direction, all methods for determining the refractive indices of LCs require a homogenous, precisely aligned sample. The optical axis can only be forced to turn in one direction by a restricted LC with a certain boundary condition at the enclosing wall and/or an externally applied field. A uniaxial crystal having an optical axis of direction pointed by a uniform nematic director, $\bf{\hat{n}}$, is analogous to a homogeneously oriented NLC. We should begin with a homogeneously oriented LC for this purpose. Additionally, a multi-wavelength Abbe refractometer that can measure refractive indices with an accuracy of $\pm 0.0002$ is the most practical and accurate tool for determining the index of refraction of an isotropic fluid. The Abbe refractometer is still the most used tool for determining the refractive indices of LCs, although it has reached two practical limits that restrict its usability: the first restriction relates to temperature control since the circulating constant temperature bath can only reach a maximum temperature of about $\sim 80$ $^{o}C$ if water is utilized, and the second arises when $n_{_{e}}$ exceeds the refractive index of glass in the refractometer ($1.3$ to $1.75$ in a low refractive index and $1.4$ to $1.85$ in a high refractive index) \cite{Mee00}.\\
\subsection{Refractive indices and nematic-isotropic first order phase transition}
Since some thermal devices are unavailable in our lab but a variety of refractive indices data is accessible for the nematic phases of rod-like LCs, we take a hypothetical LC cell as depicted in Fig. 3a into consideration to see how well our model captures the physical behavior of the LCs.
%Fig. (3)
\begin{figure}[htb]
\includegraphics[width=0.35\textwidth]{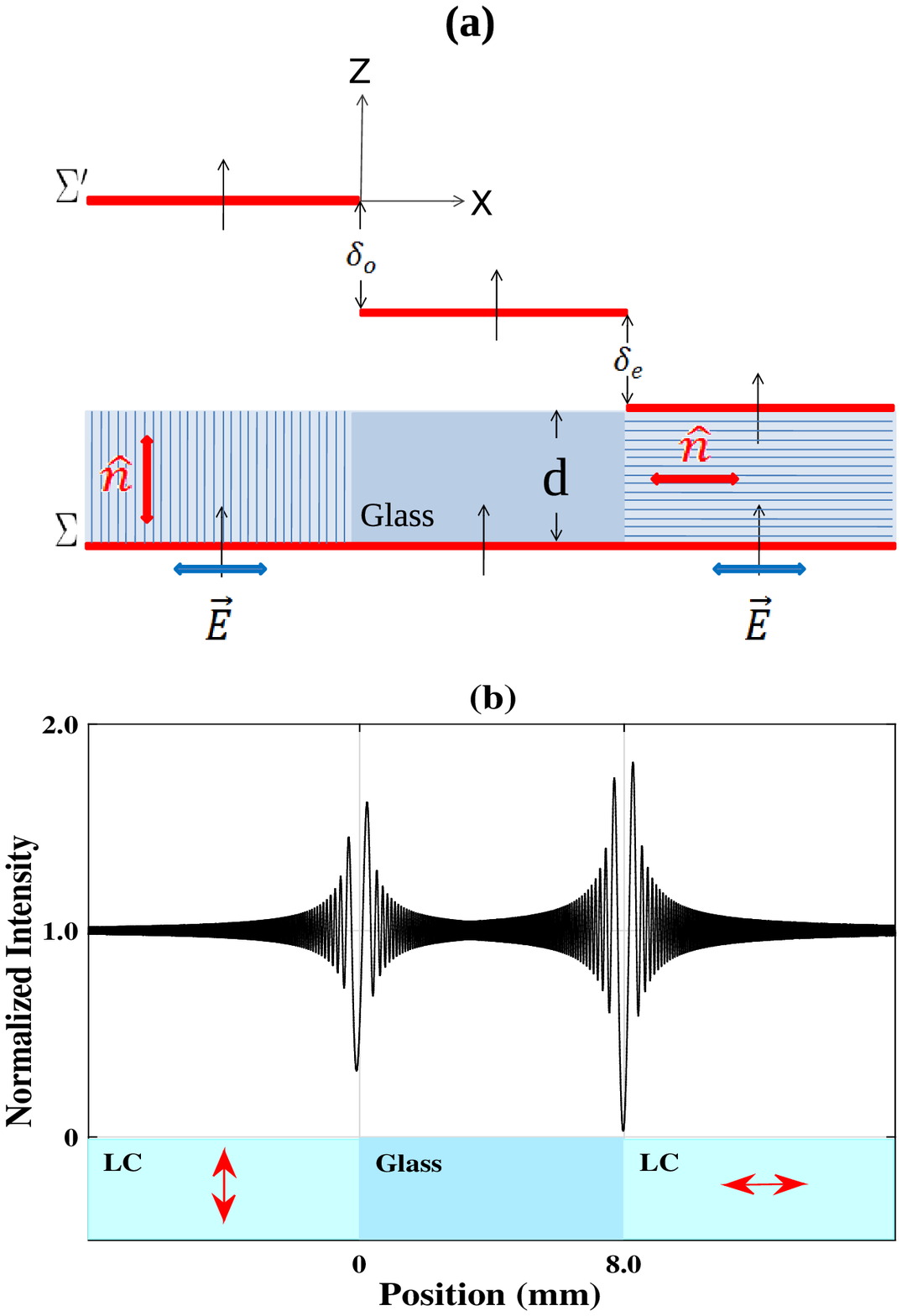}
\caption{(a) The structure of a hypothetical LC cell shows the creation of two phase steps following the passage of a coherent monochromatic plane wave-front, $\delta_{_{o}}$ and $\delta_{_{e}}$ represent the OPD corresponding ordinary and extraordinary rays, respectively. (b) Ordinary and extraordinary LCs that came into contact with the opposing sides of an $8 mm$ wide glass plate were used to calculate the normalized intensity profiles of the diffracted light from the two phase steps.}
\end{figure}
Unlike the typical LCs' cells, we used a very thin isotropic plane-parallel transparent glass plate of thickness $d$ between the two optically flat slides in the arrangement shown in Fig. 3a as substrates such that it does not completely fill the space between them.
Instead, a known uniaxial NLC can occupy the whole volume of the left and right hollow elongated rectangular cells between the two substrates made by the glass plate in order to measure the refractive indices simultaneously. Here, a coherent light wave will produce two phase steps in such a contiguous medium as the two opposite edges of the glass plate are in touch with the LC Fig. 3b. According to the simulation, the glass plate must be at least $8 mm$ wide to have a high degree of confidence that the two phase-step diffraction patterns won't overlap. The anchoring condition of the glass substrates, which creates an ordered layer at the substrate-liquid crystal interface and transmits this order to the bulk LC via elastic forces, determines the orientation of the LCs in the cell in the absence of externally applied fields. Accordingly, homogeneous anchoring in the right compartment can be achieved by unidirectionally grooving the surfaces of the two substrates parallel to one of their edges using various alignment techniques, while homeotropic anchoring in the left compartment can be achieved by using monolayer surfactants that are chemically attached to the two glass substrates. As a result, in the left and right compartments, the LC molecules are arranged parallel and perpendicular to the $z$ axis, respectively. A better alignment can also be obtained if necessary by using an external field. Additionally, consider a linearly polarized light with its electric vector in the $x$-direction. The light ray of the electric field is ordinary in the left compartment because it is parallel to the optical axis, but extraordinary in the right. It is possible to use a detector mounted on a precise computer-controlled translation stage or a linear array CCD detector at the back of the assemblage in order to record the diffraction fringes produced by the two phase steps according to their spatial extent.\\
By employing this technique, we primarily hope to identify typical LC behavior both below and beyond the thermotropic uniaxial NLC transition temperature as well as to calculate a LC's refractive indices in accordance with Eq (14). We first turn to its theoretical study to obtain a clear understanding of the visibility behavior with changing temperature and, consequently, the order parameter over a reasonably wide range of temperatures. The temperature dependent refractive indices of Eq. (9) are calculated using the known values A, B, and $\beta$ for the mesogens $4-n-pentyl-4'-cyanobiphenyl$, $C_{18}H_{19}N$ (5CB) sample \cite{LiWu04}, whose clearing point is $306.6$ $K$, and the rewritten phase differences of Eq. (10) as
%equation (15)
\begin{eqnarray}
\phi_{_{o}}=\frac{2\pi}{\lambda_{_{0}}}(n_{_{glass}}-n_{_{o}})d,\nonumber\\
           =\frac{2\pi d}{\lambda_{_{0}}}\delta n_{_{go}},\nonumber\\
\phi_{_{e}}=\frac{2\pi}{\lambda_{_{0}}}(n_{_{e}}-n_{_{glass}})d\nonumber\\
           =\frac{2\pi d}{\lambda_{_{0}}}\delta n_{_{eg}}.\nonumber\\
\end{eqnarray}
For the calculations above, we used a very light flint glass (LLF1-HIKARI glass data sheets) with a thickness of $d=10$ $\mu m$ (LLF1-HIKARI glass data sheets), whose refractive index is measured at each wavelength \cite{https}. If we assume for simplicity that both transmission coefficients are almost equal to unity, $t_{_{L}}=t_{_{R}}\cong1$, then by substituting (15) into (11), we obtain the intensity profiles of the diffracted light waves produced by the two phase steps, as shown in Fig. 3b. The ordinary and extraordinary phase differences at wavelengths $\lambda_{_{0}}= 546$, $589$ and $633$ $nm$ are numerically depicted in Figure 4 as a function of temperature in the range of $11.4$ to $54$ $^{o}C$ and with increments of $0.1$ $ ^{o}C$.
%Fig. (4)
\begin{figure}[htb]
\includegraphics[width=0.45\textwidth]{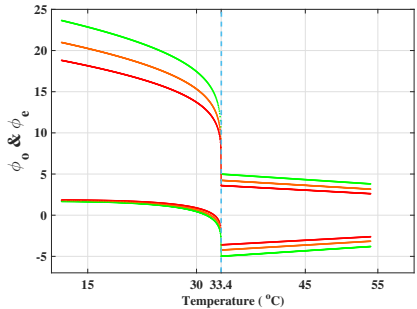}
\caption{Temperature-dependent variations in the nonlinear behavior of OPD in the nematic phase and the linear behavior in the isotropic phase for wavelengths $\lambda_{_{0}}=549$ (green), $589$ (orange), and $633$ $nm$ (red). At $T_{_{C}}=33.4^{o}C$, all these curves undergo significant change.}
\end{figure}
This figure illustrates how the phase differences in the nematic phase change nonlinearly as temperature rises and change drastically when temperature approaches the critical point at $T_{c}$. After that, the phase differences will behave linearly as the temperature rises.\\ 
We will now refer to the behavior of the visibility factor, which, as was previously said, gives us important information about what we were looking for. The visibility curves of the diffracted fringes formed by the two phase steps as functions of temperature due to the ordinary and extraordinary refractive indices of the 5CB sample are given in Figs. 5a-5d by numerical calculations using the same wavelengths, temperature range, and temperature increments as in Fig. 4.
%Fig. (5)
\begin{figure}[htb]
\includegraphics[width=0.40\textwidth]{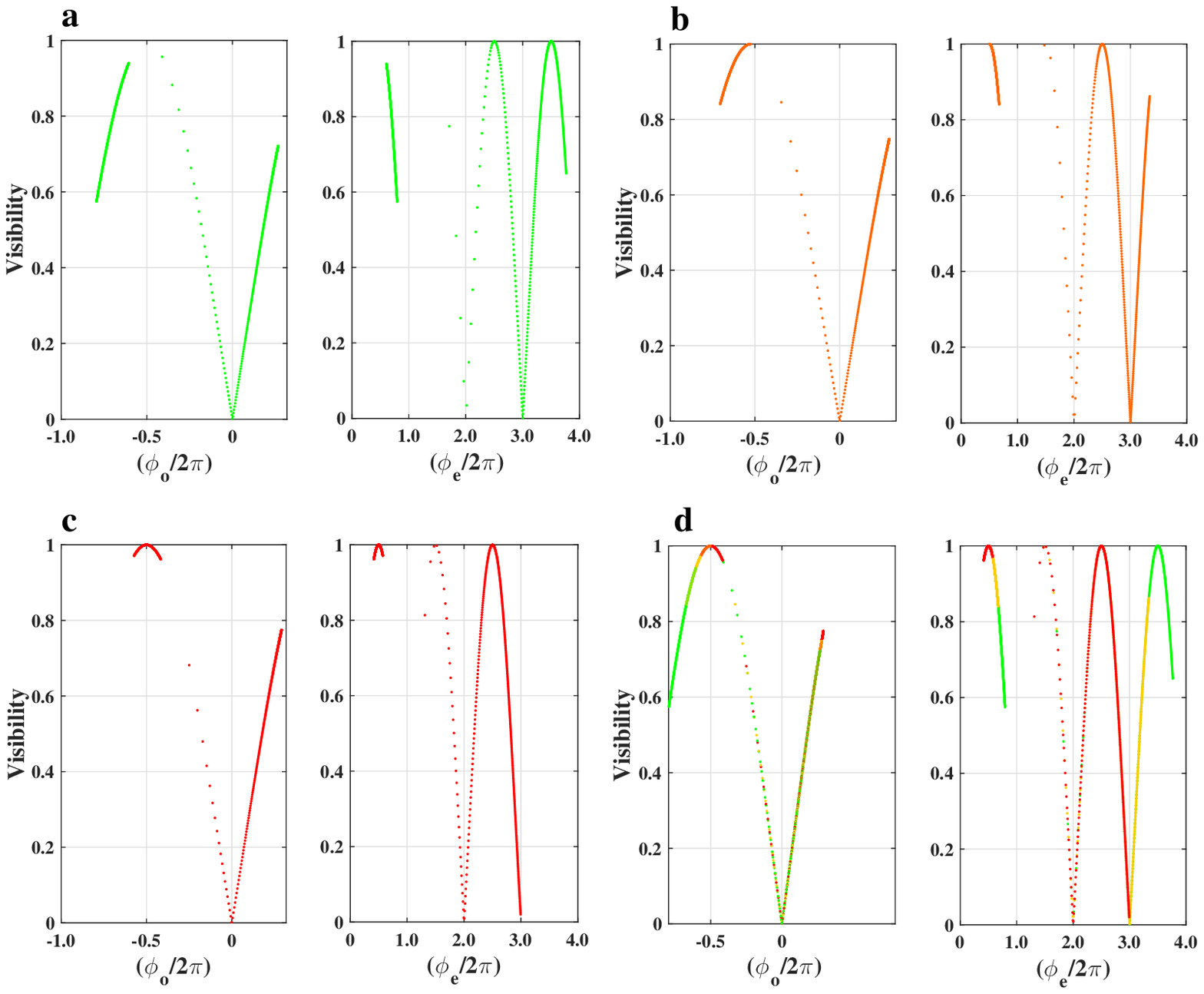}
\caption{At three wavelengths, $\lambda_{_{0}}=549, 589$, and $633$ $nm$, the visibility curves of the diffracted fringes as functions of ordinary and extraordinary normalized phase differences corresponding two phase steps have maximum value for $m=\pm 1/2, \pm 3/2, . . .$, and zero value for $m=0, \pm 1, \pm 2, . . .$.}
\end{figure}
It is evident from comparing the visibility curves of the two phase steps in the same temperature range of $11.4$ to $54$ $^{o}C$ and at a specific wavelength that the left phase step (ordinary ray) repeats its visibility curves at a considerably slower rate than the right one (extraordinary ray). This is due to the fact that in Eq. (15), $\delta n_{_{eg}} > \delta n_{_{go}}$. As long as the materials of the glass plate and the LC remain unchanged, the effect of continuous changes in temperature will therefore necessarily result in the periodic succession of visibility curves with regular sequences of maxima and zeros for any given wavelength. The visibility curves for both phase steps should change markedly theoretically and numerically as the temperature of the LC in the two compartments of the nematic phase increases gradually. Once the temperature $T_{_{C}}$ is reached, the visibility curves should abruptly change into a regime with very slow and constant changes after entering the isotropic phase, assuming that the effect of boundary conditions is ignorable as we have assumed in numerical calculations.\\
According to Figs. 6a and 6b, each of these visibility zeros exhibits an interesting situation of practical significance in optical refractometry where the diffraction fringes of one of the phase steps are entirely eliminated and a uniform intensity is predominant.
%Fig. (6)
\begin{figure}[htb]
\includegraphics[width=0.30\textwidth]{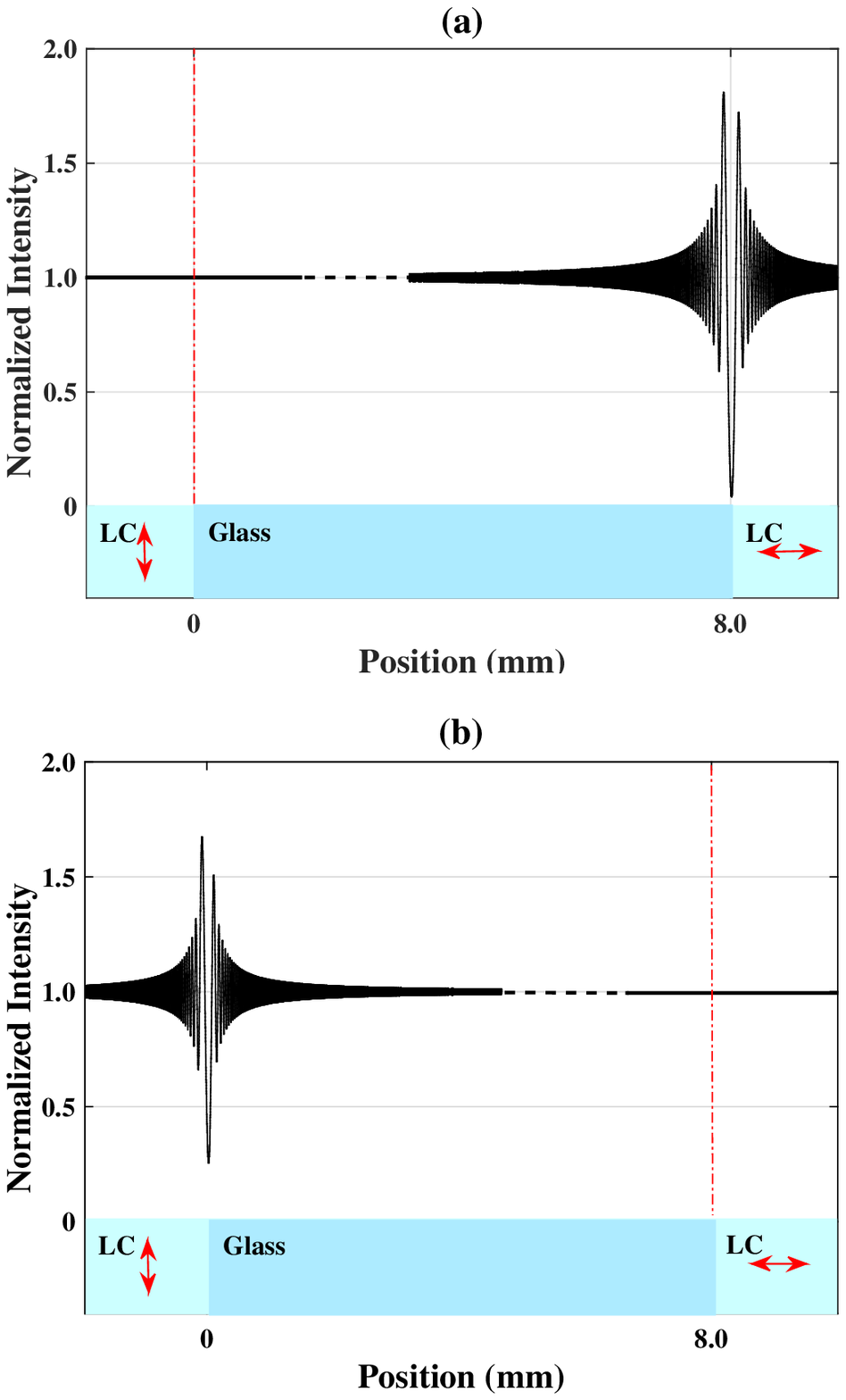}
\caption{The circumstances under which the visibility of each phase step has a zero value practically correspond to the removal of periodic fringes brought on by diffraction, which causes the light intensity to become uniform. (a) There is no diffraction effect when $\phi_{_{o}}/2\pi=0$, as the ordinary refractive index of the left LC cell is the same as that of the glass and the transmitted light will not detect any difference between the glass and the LC at a specific temperature. (b) The periodic zero condition in the Fresnel-Kirchhoff diffraction formula repeats for values of $\phi_{_{e}}/2\pi=1, 2, . . .$, causing the normalized phase difference in the right side of the LC cell to provide the precise relationship between the extraordinary refractive index and the glass as given by Eq (16).}
\end{figure}
Additionally, the visibility curves of the left and right phase steps have zero values when the following equations are true for their phase differences 
%equation (16)
\begin{eqnarray}
\phi_{_{o}}=0 \Rightarrow n_{_{o}}=n_{_{glass}},\nonumber\\
\phi_{_{e}}=2m\pi \Rightarrow n_{_{e}}=n_{_{glass}}+\frac{m\lambda_{_{0}}}{d}, m=1, 2, . . .,.
\end{eqnarray}
The formation of periodic repetition conditions in the Fresnel-Kirchhoff diffraction formula \cite{AmiTav07} is what causes the visibility curves of the right phase step to have zero values, with the exception of the value $m=0$, unlike to the left one. Naturally, for other glass plates with varying refractive indices, $m$ can generally have zero and negative values. Therefore, Eq. (16) is a second, simpler equation for calculating the values of the refractive indices in addition to Eq. (14), which is based on the fitting of experimental data. However, using a similar approach, it is also possible to calculate the refractive indices for the maximum points of visibility curves, leading to a relation that is comparable to (16). The relative simplicity, accuracy, precision, and systematic error in measurements of these two practical approaches in the laboratory all play a role in determining which of these two methods is more useful. However, despite this large benefit, the Eq. (16) has a difficulty because there are so few zeros in the curve. Fortunately, there will be a large increase in the amount of these independent data due to the availability of a wide range of optical glass plates made by optical glass producers as well as light sources with distinct wavelengths. The potential of a more thorough investigation of the fitting parameter $\beta$ — in the approximation of the order parameter proposed by Haller in Eq. (3)— and the relationship between the refractive indices and this parameter at various temperatures is made possible by these data. The most significant practical characteristic and distinctive criterion to determine whether the LC is in the isotropic phase is that the intensity patterns and visibility curves will be symmetrical, as can be seen implicitly in Figs. 5a through 5d and directly from Fig. 7. This is true for sufficiently weak or ignorable anchoring boundary conditions or at high enough temperatures in the isotropic LC.
%Fig. (7)
\begin{figure}[htb]
\includegraphics[width=0.45\textwidth]{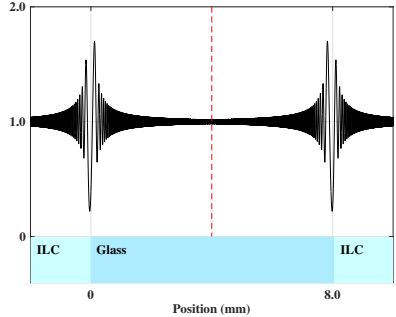}
\caption{Both the ordinary and extraordinary LCs lose their only remaining directional order and transform to the isotropic LC (ILC) for a bulk NLC at temperatures higher than the phase transition temperature $T_{_{C}}$. In this case, the light intensity patterns diffracted from the two phase steps are identical and symmetric with respect to an axis that runs parallel to the $z$ axis and through the center of the glass plate (dashed line). However, with confinement substrates, it is only conceivable to see identical and symmetrical fringes by assuming that the anchoring effect is negligible and/or that LC is at a temperature high enough to be far from $T_{_{C}}$.}
\end{figure}
As a result, as soon as this important requirement is observed, it is clear that the uniaxial NLC is in the isotropic phase $(Q=S=0)$, which has the highest degree of symmetry, and is far from both the weak and strong anchoring constraints.\\
Utilizing the numerical results of (15), (11) and (12), the visibility curves in Fig. 5 were made possible. With a precise step of the temperature increment that may be made as small as practically possible and recording the visibility of fringes at any temperature, the same process may be achieved in the lab in practice. Unlike the numerical method, the number of data in this method is constrained by the material's operating temperature range and the sensitivity of the thermal instruments available in the lab. If we fit the simulated visibility curve to the obtained data for each period of repeated visibility patterns, then by (14) it is possible to derive refractive indices with excellent agreement. Ordinary and extraordinary refractive indices of the 5CB sample are shown in Figure 8 as functions of temperature, with three colored circles representing each wavelength. 
%Fig. (8)
\begin{figure}[htb]
\includegraphics[width=0.45\textwidth]{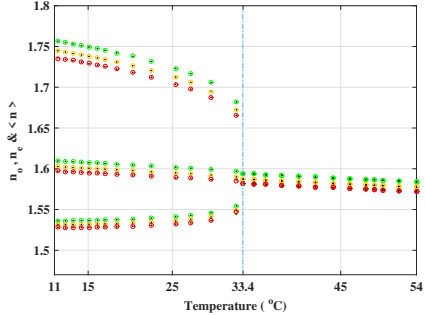}
\caption{Extraction of $n_{_{o}}$, $n_{_{e}}$ and $< n >=(n_{_{e}}+2n_{_{o}})/3$ in the two nematic and isotropic phases using Eq. 14 for three wavelengths to fit on experimentally measured visibility curves obtained from sample 5CB. We have merged the experimental data (black dots) and the calculations from Eq. (14) (colored circles) into one graph in order to compare their high degree of congruence.} 
\end{figure}
In order to compare the method with the results of the numerical analysis, experimental data with black dots are provided \cite{LiWu04}. Additionally, the average refractive index of LC, which exhibits linear behavior with respect to temperature, is observable at three separate wavelengths in both the nematic and isotropic phases. It is evident that the numerical refractive indices obtained are in excellent agreement with their experimental values, demonstrating the efficacy and precision of this procedure. At $T_{_{C}}$, where these two refractive indices become singular, this figure also includes the most significant practical application of the phase transition of the first order from nematic phase to isotropic phase.\\
Despite the similarities in the patterns of the visibility curves for the two phase steps at various wavelengths, particularly at their maxima and zeros, there are two crucial differences. The first difference comes at the start and finish of the curves for different wavelengths, as is immediately seen from the visibility curves. However, there is another distinction that is less obvious and is only detectable by numerical computations. The output of Eq. (16)'s numerical computations for the ordinary and extraordinary refractive indices at the visibility zeros is shown in Table 1.
%Table. 1
\begin{table}[htb]
\caption{The ordinary and extraordinary refractive indices of a 5CB sample at three different wavelengths at various temperatures when the visibilities have theoretically fallen to zero}
\centering 
\begin{tabular}{c c c c}
\hline\hline
$\lambda$&$\frac{\phi_{_{o}}}{2\pi}=0$&$\frac{\phi_{_{e}}}{2\pi}=2$&$\frac{\phi_{_{e}}}{2\pi}=3$\\
$(\mu m)$&$n_{_{o}}=n_{_{glass}}$&$n_{_{e}}=n_{_{glass}}+\frac{2\lambda_{_{0}}}{d}$&$n_{_{e}}=n_{_{glass}}+\frac{3\lambda_{_{0}}}{d}$\\[0.7ex]
\hline\\
0.546&$31.34^{o}C$&$33.16^{o}C$&$27.61^{o}C$\\
&$n_{_{o}}=1.5510$&$n_{_{e}}=1.6602$&$n_{_{e}}=1.7148$\\\\
0.589&$31.91^{o}C$&$32.63^{o}C$&$20.92^{o}C$\\
&$n_{_{o}}=1.5481$&$n_{_{e}}=1.6659$&$n_{_{e}}=1.7248$\\\\
0.633&$32.44^{o}C$&$31.57^{o}C$&$11.13^{o}C$\\
&$n_{_{o}}=1.5456$&$n_{_{e}}=1.6722$&$n_{_{e}}=1.7355$\\\\
\hline\hline
\end{tabular}
\end{table}\\
The first-order nematic-isotropic phase transition's impact on these diagrams is not entirely evident, despite the fact that we are somewhat familiar with how visibility curves behave around phase transitions. A more striking criterion around the phase transition temperature $T_{_{C}}$ will be obtained if we look at the visibility gradient. In terms of temperature, Fig. 9 shows the typical and exceptional visibility gradient behaviors for the wavelength of $633$ $nm$, which are essentially the same for other wavelengths.
%Fig. (9)
\begin{figure}[htb]
\includegraphics[width=0.45\textwidth]{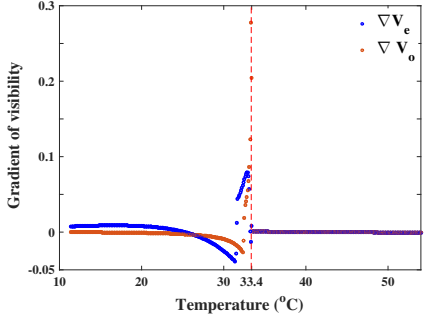}
\caption{Visibility gradients behavior at wavelength $\lambda_{_{0}}=633$ $nm$, which are typically true for other wavelengths as well. Using the visibility gradients, it is possible to obtain three distinctly different behaviors with temperature: nonlinear in the nematic phase, constant in the isotropic phase, and rapid and significant variations around $T_{_{C}}$, which indicate the singularity in this temperature. The visibility gradient diagrams will lead to the same conclusions, which we may infer from Figs. 4 and 5, clearly evident.}
\end{figure}
As we previously discussed, it is easy to notice in this figure how the ordinary and extraordinary visibility gradients of LC in the nematic phase, before reaching the phase transition temperature, behave in distinct ways. We observe a behavior with zero gradient and strikingly similar regime of ordinary and extraordinary visibility in the isotropic phase, in contrast to the distinct behavior before the phase transition temperature and quite remarkable at the phase transition temperature, which indicates a singularity in the visibility gradient. It is deserving of investigation and in-depth physical analysis that the visibility gradient undergoes abrupt changes around the phase transition temperature and somehow reflects the pretransition circumstances. Although numerical computations with a very small temperature increment allow for a more in-depth analysis, they are neither practicable for use in the lab nor likely to have a physical counterpart when used as artificial data. However, in this study, we are just examining the viability of the current approach in dealing with the first and second order phase transition phenomena. \\
\subsection{First and second order phase transitions in a surface-aligned nematic film}
As already indicated, unlike the simple relationship (3), the order parameter $S$ of the NLC molecules is mostly the result of two conflicting thermal fluctuations, intermolecular interactions, and/or external fields. Without an external field, the alignment agent results from the interaction of the LC molecules with substrate regularities. As a result of this intermolecular interaction, the alignment agent extends to points far from and normal to the substrate-liquid crystal interface up to a long range, which is determined by the correlation length. As a result, as the temperature rises, the thermal fluctuation inhibits the alignment agent and results in a reduction in the correlation length and order parameter. It was noted in the previous section that the first-order nematic-isotropic phase transition process in the bulk of LCs not only manifests itself as a singularity in the refractive indices but also in the visibility gradient, and it was for this reason that these were introduced as criteria for identifying the first-order phase transition. It is intended to use this method and its capabilities to identify nematic-isotropic phase transitions in a surface-aligned nematic film using a simple model that is not too distant from reality. It is demonstrated there that, if the anchoring of the LC to the substrates were disregarded, the visibility of the fringes of the two phase steps in Fig. 7 would be symmetric as soon as the phase transition temperature $T_{_{C}}$ from nematic to isotropic was passed. As we shall see, the refractive indices and the gradient of visibilities in the nematic phase and in the immediate vicinity of the first-order phase transition temperature will still change significantly despite the fact that one cannot realistically expect any symmetry in the curves of the two visibility curves in the presence of confinement substrates. As the isotropic phase transition at $T_{_{C}}$ approaches, the long-range orientational order in the bulk of NLCs decreases to a very short-range one, and the long-range correlation length eventually approaches zero. But according to Fig. 10, we can assume that the two aligned and uniform nematic films of LC with thicknesses $D_{o}$ and $D_{e}$ adherent to the upper and lower substrates, respectively, are created in the left and right compartments, despite the nematic-isotropic phase transition in the bulk of the LC and due to the dominance of the order agent in the vicinity of the liquid crystal-substrate boundary.
%Fig. (10)
\begin{figure}[htb]
\includegraphics[width=0.45\textwidth]{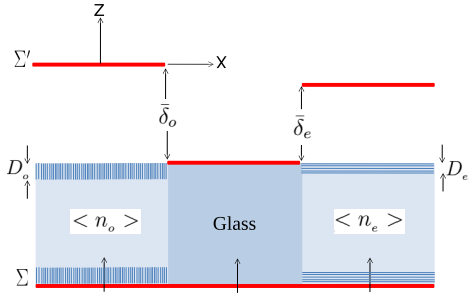}
\caption{After raising the temperature from the first-order phase transition in the bulk of LC, surface-aligned nematic films with thicknesses $D_{_{o}}$ and $D_{_{e}}$ attached to the substrates, respectively, form on the left and right sides of the LC cell, causing the average refractive indices to be $<n_{_{o}} >$ and $< n_{_{e}} >$ in these two portions.}
\end{figure}
Using the Landau-de Gennes theory, there is a critical film thickness of $D^c$, and if the newly produced layer is thicker than that (i.e. $D_{o, e}>D^c_{o, e}$), the desired film will go through the first-order phase transition when the temperature is raised above $T_{_{C}}$ \cite{She76}. If not, a second-order phase transition will occur. The following relation could be used to mathematically express this
%equation (17)
\begin{equation}
T^{S}_{o,e}=T_{_{C}}+\delta T_{o,e},
\end{equation}
where $T^{S}_{o,e}$, $T_{_{C}}$, and $\delta T_{o,e}$ ($\delta T_{o,e}>0$) are respectively, the ordinary and extraordinary surface-aligned nematic films' phase transition temperatures, the first-order nematic-isotropic phase transition in the bulk of LC and their respective temperature increments. The amount of $\delta T_{o,e}$ and, consequently, the difference between the phase transition in the bulk and the phase transition in the surface-aligned nematic film will rise when the ordering agent is increased or strong anchoring is obtained. To acquire reliable data, it is required to increase the value of $\delta T_{o,e}$ in accordance with the experimentally measured values of $\delta T_{o,e}$, which are typically a few tenths of kelvin. First, using the suggested model, we look into the condition that leads to the first-order phase transition in the surface-aligned nematic film (i.e. $D_{o, e}>D^c_{o, e}$). The phase transition will inevitably be of the second order if the prerequisites for producing this transition are not met in practice. By measuring the spatial averaged refractive index over a length interval of $d$ or equal optical length of a beam travelling through the left and right compartments, we then obtain \cite{BorWol99}
%equation (18)
\begin{eqnarray}
<n_{_{o}}(T)>=\frac{n_{_{i}}(T)(d-2D_{_{o}}(T))+2n_{_{o}}(T)D_{_{o}}(T)}{d},\nonumber\\
\simeq n_{_{i}}(T)-0.1[n_{_{i}}(T)-n_{_{o_{min}}}]\nonumber\\
<n_{_{e}}(T)>=\frac{n_{_{i}}(T)(d-2D_{_{e}}(T))+2n_{_{e}}(T)D_{_{e}}(T)}{d},\nonumber\\
\simeq n_{_{i}}(T)+0.1[n_{_{e_{max}}}-n_{_{i}}(T)],
\end{eqnarray}
where $n_{_{i}}(T)$ is the isotropic phase's bulk refractive index of LC at temperature $T$. Additionally, by forming the surface-aligned nematic films in contact with the substrate boundaries and the bulk of the LC, and assuming that $\delta T_{o}\simeq\delta T_{e}$, the average optical birefringence at a specific wavelength denoted as $\bar{\Delta n}$ then becomes
%equation (19)
\begin{eqnarray}
\bar{\Delta n}=<n_{_{e}}>-<n_{_{o}}>,\nonumber\\
=0.1(n_{_{e_{max}}}-n_{_{o_{min}}}).
\end{eqnarray}
For example, in terms of practical estimation, within the limits of acceptable laboratory approximations
%equation (*)
\begin{eqnarray}
\begin{cases}
\delta T_{o,e}\simeq 1K\nonumber\\
n_{_{o}}(T)\simeq n_{_{o_{min}}}\nonumber\\
n_{_{e}}(T)\simeq n_{_{e_{max}}},\nonumber\\
\frac{2D_{_{e}}(T)}{d}\simeq 0.1,
\end{cases}
\end{eqnarray}
This is practically measurable and might cause a shift in the relative refractive index of roughly $10^{-2}$.
By rewriting relations (15) and use the average refractive indices in Eqs. (18), we will have
%equation (20)
\begin{eqnarray}
\bar{\phi_{_{o}}}=\frac{2\pi}{\lambda_{_{0}}}(n_{_{glass}}-<n_{_{o}}>)d,\nonumber\\
           =\frac{2\pi d}{\lambda_{_{0}}}\delta \bar{n_{_{go}}},\nonumber\\
\bar{\phi_{_{e}}}=\frac{2\pi}{\lambda_{_{0}}}(<n_{_{e}}>-n_{_{glass}})d\nonumber\\
           =\frac{2\pi d}{\lambda_{_{0}}}\delta \bar{n_{_{eg}}}.
\end{eqnarray}
Based on the foregoing hypotheses, behavior of the refractive indices as a function of temperature in three cases - the nematic phase until the first-order phase transition in the bulk, $T_{_{C}}$, the first-order phase transition in the surface-aligned nematic film, $T^{S}_{o,e}$, and finally the homogeneous isotropic phase in all points of the LC - is calculated numerically and shown in Fig. 11.\\
%Fig. (11)
\begin{figure}[htb]
\includegraphics[width=0.45\textwidth]{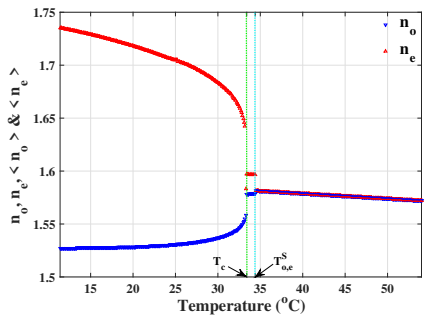}
\caption{Variations for $n_{_{o}}$ and $n_{_{e}}$ in the bulk of LC during the nematic phase, for $< n_{_{o}} >$ and $< n_{_{e}} >$ resulting from their regular development in the surface-aligned nematic films, and lastly for the bulk of the LC during the isotropic phase. The average optical birefringence of anisotropic LC, $\bar{\Delta n}$, occurs in the temperature range between the first-order transitions of the LC bulk in the nematic phase and the LC bulk in the isotropic phase, $T_{_{C}} < T < T^{S}_{_{o,e}}$, after the development of the ordered surface layer.}
\end{figure}
According to Fig. 11, the anchoring effect will result in the formation of regular surface film layers on the left and right sides of the LC cell, which will change the behavior of the refractive indices $n_{_{o}}=n_{_{e}}=n_{_{i}}$ in the isotropic phase in a way that differs from the ideal state depicted in Fig. 8 and result in measurable changes to the average refractive indices $< n_{_{o}} >$ and $< n_{_{e}} >$ ($< n_{_{o}} > \neq < n_{_{e}} > \neq n_{_{i}}$). As a result, after raising the temperature above $T_{_{C}}$, the average optical birefringence of anisotropic LC, $\bar{\Delta n}$, will be formed. As opposed to the nematic phase state, where the light outputs from both sides of the LC cell have extremely high degrees of linear polarization, it is also anticipated that in this situation the output light will have relatively low degrees of partial polarization. Additionally, using default settings and numerical calculations, it can be seen that the visibility gradients will change when regular surface films are taken into account; this change is much more noticeable in the extraordinary gradient visibility than the ordinary gradient visibility, as shown in Fig. 12.\\
%Fig. (12)
\begin{figure}[htb]
\includegraphics[width=0.50\textwidth]{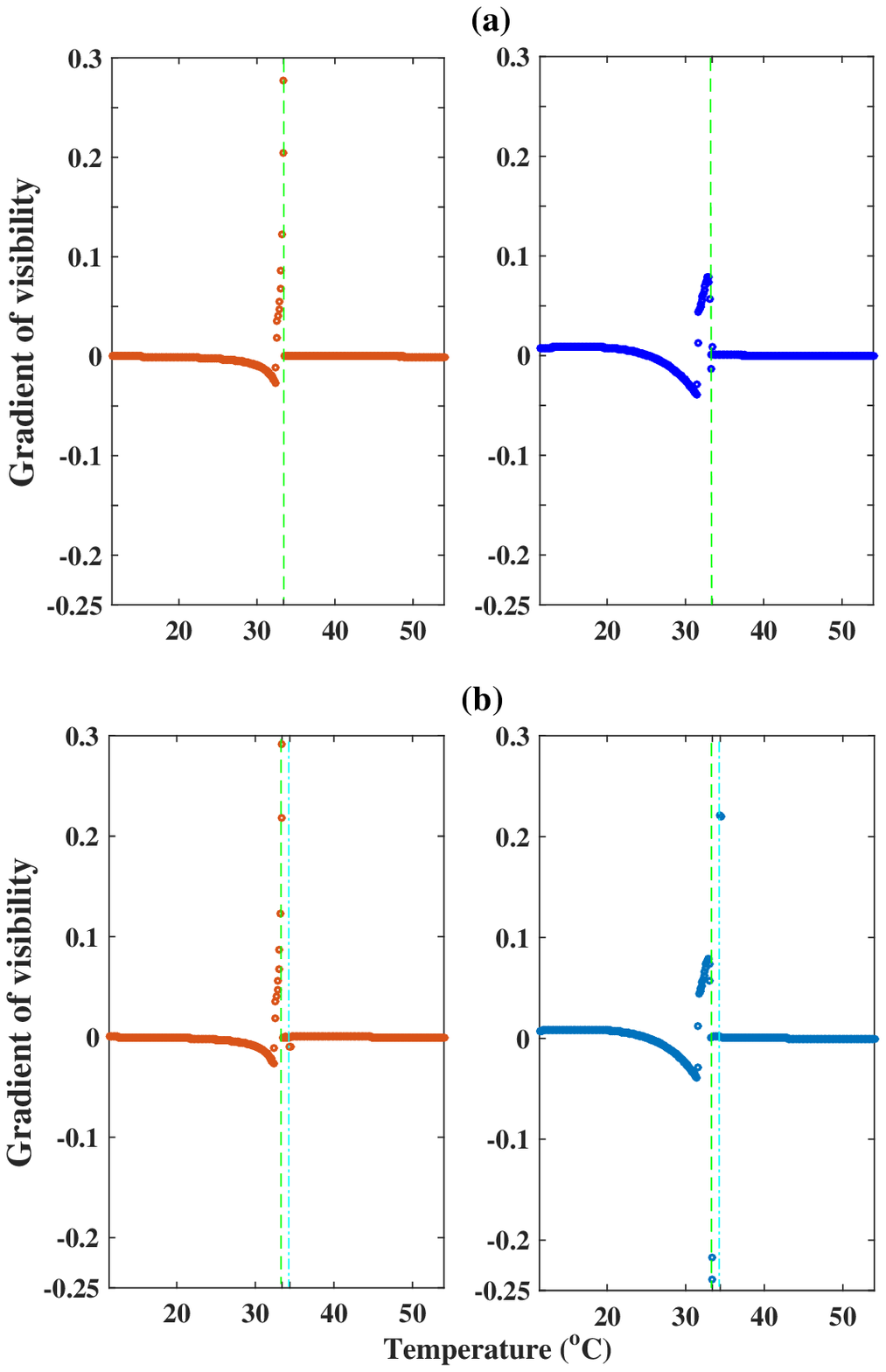}
\caption{Comparing the gradient visibilities: (a) Without taking into account the ordered surface film effect. (b) By taking into account the ordered surface film effect. It is evident that whereas the ordinary visibility gradient has barely changed, the extraordinary visibility gradient has seen major modifications.}
\end{figure}
The utilization of temperature gradient in the bulk of LCs is also an intriguing illustration of the multiple applications of FD from phase steps. In this technique, a desired LC is brought into contact with two separate heat reservoirs through a fully controlled process at two different temperatures, creating a significant directional component of temperature gradient (say along the x axis). For instance, the phase transition sequence for $4-4'-dipentylbiphenil$ is as follows \cite{Czu94}:\\
$C$ 25.1 $S_{E}$ 46.1 $S_{E'}$ 47.1 $S_{B}$ 52.3 $I$.\\
When an ordinary ray of a coherent monochromatic light wave propagating along the $z$ axis encounters the separation boundary of each of the two adjacent phases, provided that each phase is sufficiently transparent, diffraction will occur if the system is placed in a suitable directional temperature gradient and in a steady-state regime. In this example, the total number of phases that are created in an appropriate temperature gradient, whose interval and rate of change can be altered, will be discovered all at once and in a single shot. The assumption of approximate uniformity of the temperature gradient applied to the LC, which is probably not always compatible with the temperature behavior in different phases, can also be used to estimate the temperature range of the phase transition between separate phases.
% 
%-------------------------------------------------------------------------------------------------------------------
%
\section*{Conclusions}
Since researchers began taking a serious and methodical approach to the study of LCs because of their diversity of phases and numerous properties, these materials have benefited from a privileged position because of their many and expanding applications in current technology as well as their potentials in the horizons of the future. Because LC phase changes are formed at different temperatures as a result of heat exchange between the LC and its environment, the measurement of parameters related to free energy and heat flow is regarded to be significant in theoretical and experimental studies. Given that light is sensitive to changes within LCs and that FD from phase objects has significantly contributed to precise optical metrology over the past two decades, the current theoretical research is attempting to accomplish two major achievements of its potentials in theoretical investigations of LCs. The basis of this method is the utilization of the visibility parameter of the three central diffracted light fringes, which after interaction with a singularity created during the passage of a coherent light wave-front and has great accuracy due to their symmetry, periodicity, mathematical and analytical simplicity. The first achievement is the accurate measurement of both the ordinary and extraordinary refractive indices at the same time, as well as the removal of the limitation and disadvantage of using the traditional Abbe refractometer to determine the extraordinary refractive index of LCs. Additionally, it is feasible to validate the theoretical models relating to the order parameter and microscopic polarizabilities using the results.\\
The use of ordinary and extraordinary visibility gradient diagrams, which have different and distinct behavior in the nematic phase, similar behavior and zero value in the isotropic phase, and significant and rapid change around the phase transition temperature, is the second achievement in addition to the discrete and continuous behavior of the refractive indices in the first and second order phase transitions, respectively. In the case of phase transitions, such as those with low enthalpy changes, the advantages of the present method can clearly overcome the drawbacks and shortcomings of conventional methods. In the end, this model, like any simplistic model at first, undoubtedly overlooks some crucial quantities, but by improving the model and adapting it to the problem's useful factors, it demonstrates its effectiveness. Therefore, it may be hoped that the new way will open a door in the face of current difficulties and in response to the shortcomings of the traditional approaches, and that it will be proposed as a fresh strategy alongside the effective existing methods.
\newpage
\section*{List of Figure Captions}
Fig. 1: (a) The optical length of a direct ray at normal incidence is shown to illustrate the production of a phase step after passing a coherent monochromatic plane wave-front from two adjacent phase objects with different refractive indices. (b) A simulation of the intensity outlook determined at a distance of $R=5$ $cm$ from the back of the glass plate. (c) Calculated normalized intensity distribution of diffracted light along the yellow dashed line, corresponding to (b).

Fig. 2: Fringe visibility variation with normalized OPD, $(\delta/\lambda_{_{0}})=\phi/2\pi$. For both temperature-dependent linear and non-linear variation of OPD, this function has a periodic and symmetrical structure with an axis of symmetry that has reached its maximum at values of $\phi/2\pi=\pm 1/2, \pm 3/2, . . .$ and reaches zero at values of $\phi/2\pi =0, \pm 1, \pm 2, . . .$.

Fig. 3: (a) The structure of a hypothetical LC cell shows the creation of two phase steps following the passage of a coherent monochromatic plane wave-front, $\delta_{_{o}}$ and $\delta_{_{e}}$ represent the OPD corresponding ordinary and extraordinary rays, respectively. (b) Ordinary and extraordinary LCs that came into contact with the opposing sides of an $8 mm$ wide glass plate were used to calculate the normalized intensity profiles of the diffracted light from the two phase steps.

Fig. 4: Temperature-dependent variations in the nonlinear behavior of OPD in the nematic phase and the linear behavior in the isotropic phase for wavelengths $\lambda_{_{0}}=549$ (green), $589$ (orange), and $633$ $nm$ (red). At $T_{_{C}}=33.4^{o}C$, all these curves undergo significant change.

Fig. 5: At three wavelengths, $\lambda_{_{0}}=549, 589$, and $633$ $nm$, the visibility curves of the diffracted fringes as functions of ordinary and extraordinary normalized phase differences corresponding two phase steps have maximum value for $m=\pm 1/2, \pm 3/2, . . .$, and zero value for $m=0, \pm 1, \pm 2, . . .$. 

Fig. 6: The circumstances under which the visibility of each phase step has a zero value practically correspond to the removal of periodic fringes brought on by diffraction, which causes the light intensity to become uniform. (a) There is no diffraction effect when $\phi_{_{o}}/2\pi=0$, as the ordinary refractive index of the left LC cell is the same as that of the glass and the transmitted light will not detect any difference between the glass and the LC at a specific temperature. (b) The periodic zero condition in the Fresnel-Kirchhoff diffraction formula repeats for values of $\phi_{_{e}}/2\pi=1, 2, . . .$, causing the normalized phase difference in the right side of the LC cell to provide the precise relationship between the extraordinary refractive index and the glass as given by Eq (16).

Fig 7: Both the ordinary and extraordinary LCs lose their only remaining directional order and transform to the isotropic LC (ILC) for a bulk NLC at temperatures higher than the phase transition temperature $T_{_{C}}$. In this case, the light intensity patterns diffracted from the two phase steps are identical and symmetric with respect to an axis that runs parallel to the $z$ axis and through the center of the glass plate (dashed line). However, with confinement substrates, it is only conceivable to see identical and symmetrical fringes by assuming that the anchoring effect is negligible and/or that LC is at a temperature high enough to be far from $T_{_{C}}$.

Fig. 8: Extraction of $n_{_{o}}$, $n_{_{e}}$ and $< n >=(n_{_{e}}+2n_{_{o}})/3$ in the two nematic and isotropic phases using Eq. 14 for three wavelengths to fit on experimentally measured visibility curves obtained from sample 5CB. We have merged the experimental data (black dots) and the calculations from Eq. (14) (colored circles) into one graph in order to compare their high degree of congruence.

Fig. 9: Visibility gradients behavior at wavelength $\lambda_{_{0}}=633$ $nm$, which are typically true for other wavelengths as well. Using the visibility gradients, it is possible to obtain three distinctly different behaviors with temperature: nonlinear in the nematic phase, constant in the isotropic phase, and rapid and significant variations around $T_{_{C}}$, which indicate the singularity in this temperature. The visibility gradient diagrams will lead to the same conclusions, which we may infer from Figs. 4 and 5, clearly evident.

Fig. 10: After raising the temperature from the first-order phase transition in the bulk of LC, surface-aligned nematic films with thicknesses $D_{_{o}}$ and $D_{_{e}}$ attached to the substrates, respectively, form on the left and right sides of the LC cell, causing the average refractive indices to be $<n_{_{o}} >$ and $< n_{_{e}} >$ in these two portions. 

Fig. 11: Variations for $n_{_{o}}$ and $n_{_{e}}$ in the bulk of LC during the nematic phase, for $< n_{_{o}} >$ and $< n_{_{e}} >$ resulting from their regular development in the surface-aligned nematic films, and lastly for the bulk of the LC during the isotropic phase. The average optical birefringence of anisotropic LC, $\bar{\Delta n}$, occurs in the temperature range between the first-order transitions of the LC bulk in the nematic phase and the LC bulk in the isotropic phase, $T_{_{C}} < T < T^{S}_{_{o,e}}$, after the development of the ordered surface layer.

Fig. 12: Comparing the gradient visibilities: (a) Without taking into account the ordered surface film effect. (b) By taking into account the ordered surface film effect. It is evident that whereas the ordinary visibility gradient has barely changed, the extraordinary visibility gradient has seen major modifications.
%
%---------------------------------------------------------------------------------------------------------------------
%
\newpage
\section*{List of Table Caption}

Table 1: The ordinary and extraordinary refractive indices of a 5CB sample at three different wavelengths at various temperatures when the visibilities have theoretically fallen to zero
%
%---------------------------------------------------------------------------------------------------------------------
%
\end{document}